\pdfoutput=1
\documentclass[12pt,letterpaper,titlepage,reqno,oneside]{article}
\usepackage[utf8]{inputenc}
\usepackage{doi}
\usepackage{graphicx}
\usepackage[super,comma]{natbib}
\usepackage[english]{babel}
\usepackage[figurename=FIG.,tablename=TABLE]{caption}
\usepackage{enumerate}
\usepackage{amsmath,amsfonts,amssymb,wasysym,latexsym,amsthm,bm}
\usepackage[mathscr]{eucal}
\usepackage{color}
\usepackage{subfig}
\usepackage{siunitx}
\usepackage{textcomp}
\usepackage{titlesec}
\usepackage{multirow}
\usepackage[subfigure,titles]{tocloft}
\usepackage{setspace}

\DeclareCaptionFormat{upper}{#1#2\uppercase{#3}\par}

\titleformat{\section}{\normalsize\bfseries}{\Roman{section}.}{1em}{\normalsize\uppercase}
\titleformat{\subsection}{\normalsize\bfseries}{\Alph{subsection}.}{1em}{}
\titleformat{\subsubsection}{\normalsize\itshape}{\arabic{subsubsection}.}{1em}{}

\renewcommand{\thesection}{\Roman{section}}

\cftpagenumbersoff{figure}
\cftpagenumbersoff{subfigure}

\makeatletter
\renewcommand\listoffigures{%
    \medskip\raggedright \textbf{LIST OF FIGURES}%
    \@mkboth{\MakeUppercase\listfigurename}%
        {\MakeUppercase\listfigurename}%
    \@starttoc{lof}%
}
\makeatother

\makeatletter
\def\p@subsection{\thesection.\,}
\makeatother

\begin{document}
\sisetup{detect-family,detect-display-math=true,exponent-product=\cdot,output-complex-root=\text{\ensuremath{i}},per-mode=symbol}

\author{%
Michael A.G. Timmer\footnote{Author to whom correspondence should be addressed. E-mail: m.a.g.timmer@utwente.nl}, Joris P. Oosterhuis, Simon B\"uhler, \\ and Theo H. van der Meer \\
\textit{Department of Thermal Engineering, University of Twente,}\\
\textit{Enschede, The Netherlands}\\ \\
Douglas Wilcox\\
\textit{Chart Inc., Troy, New York 12180}%
}
\title{\LARGE{Characterization and reduction of flow separation in jet pumps for laminar oscillatory flows} \\
\textit{\large{Reducing flow separation in jet pumps}}
}
\date{\today}

\maketitle

\begin{abstract}
A computational fluid dynamics model is used to predict the oscillatory flow through tapered cylindrical tube sections (jet pumps). The asymmetric shape of jet pumps results in a time-averaged pressure drop that can be used to suppress Gedeon streaming in closed-loop thermoacoustic devices. However, previous work has shown that flow separation in the diverging flow direction counteracts the time-averaged pressure drop. In this work, the characteristics of flow separation in jet pumps are identified and coupled with the observed jet pump performance. Furthermore, it is shown that the onset of flow separation can be shifted to larger displacement amplitudes by designs that have a smoother transition between the small opening and the tapered surface of the jet pump. These design alterations also reduce the duration of separated flow, resulting in more effective and robust jet pumps. To make the proposed jet pump designs more compact without reducing their performance, the minimum big opening radius that can be implemented before the local minor losses have an influence on the jet pump performance is investigated. To validate the numerical results, they are compared with experimental results for one of the proposed jet pump designs. \\

\end{abstract}

\addtocounter{page}{2}

\section{Introduction}
\label{sec:intro}
A promising kind of thermoacoustic devices is based on a traveling wave configuration consisting of a closed-loop tube. \cite{Ceperly1979,Backhaus1999} Despite the high achievable efficiency, one of the main problems of these designs is the possible occurrence of a time-averaged mass flux (Gedeon streaming). \cite{Gedeon1997b} In a closed-loop configuration, this type of acoustic streaming causes unwanted convective heat transport that has a detrimental effect on the device's efficiency. Control of Gedeon streaming is therefore crucial to achieve efficient closed-loop thermoacoustic devices.\cite{Swift1999}

\begin{figure}
	\centering
	\includegraphics[width=.5\textwidth]{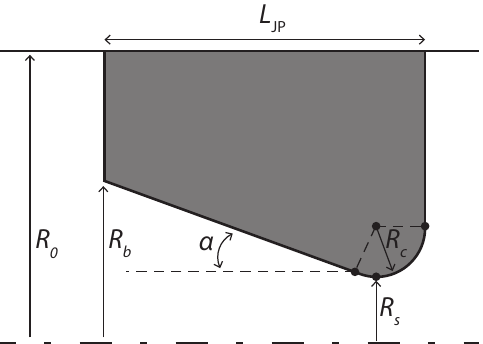}
	\caption{Axisymmetric representation of a jet pump with parameters that define the geometry. Bottom dashed line indicates center line, top solid line indicates outer tube wall. Reproduced from~\cite{Oosterhuis2015}.}
	\label{fig:jetpumpgeom}
\end{figure}

A frequently used manner to suppress Gedeon streaming is the utilization of a jet pump. \cite{Backhaus2000,Biwa2007,Boluriaan2003,Boluriaan2003a} This is a static component consisting of a tapered tube section with two unequally sized openings. The asymmetric design causes an imbalance in the hydrodynamic end effects, resulting in a time-averaged pressure drop across the jet pump. By balancing the time-averaged pressure drop of the jet pump with that caused by the rest of the thermoacoustic device, Gedeon streaming can be suppressed. \cite{Backhaus2000} A schematic of a typical jet pump geometry with its parameters is depicted in Fig. \ref{fig:jetpumpgeom}. The outer tube has a radius of $R_0$ and the big opening and small opening (``waist") have a radius of $R_b$ and $R_s$, respectively. A curvature of $R_c$ is applied at the jet pump waist to decrease the local contraction minor losses, therewith increasing the time-averaged pressure drop. Given the aforementioned parameters, the tapered surface with angle $\alpha$ determines the axial jet pump length $L_{\text{JP}}$, or vice versa.

\subsection{Quasi-steady approximation}
The current design guidelines for jet pumps are based on a quasi-steady approximation derived by Backhaus and Swift. \cite{Backhaus2000} The jet pump performance is approximated by using minor loss coefficients reported for the abrupt expansion and contraction in steady pipe flow. Following Ohmi and Iguchi, it is assumed that the oscillatory flow can be decomposed in two quasi-steady flows. \cite{Iguchi1982} The jet pump performance in terms of the time-averaged pressure drop is then given by, \cite{Backhaus2000} 
\begin{align}
	\Delta p_{\text{2,JP}} & = \frac{1}{8} \rho_0 |u_{\text{1,JP}}|^2  \left[(K_{\text{exp},s}-K_{\text{con},s}) + \left(\frac{A_s}{A_b}\right)^2 (K_{\text{con},b}-K_{\text{exp},b}) \right], 
	\label{eq:dp2}
\end{align}  
where $|u_{\text{1,JP}}|$ is the radially averaged velocity amplitude inside the jet pump waist and the subscripts ``$s$" and ``$b$" denote the small opening and the big opening, respectively. $K_{\text{exp}}$ is the minor loss coefficient for expanding flow and $K_{\text{con}}$ is the minor loss coefficient for contracting flow, which are both well documented for a wide range of geometries in steady flows. \cite{Idelchik2007} 

The penalty of utilizing minor losses to establish a time-averaged pressure drop is the associated dissipation of acoustic power. Following the aforementioned quasi-steady approximation, the jet pump performance in terms of the time-averaged acoustic power dissipation is given by, \cite{Backhaus2000}
\begin{align}
\Delta \dot{E}_{\text{2,JP}} & = \frac{1}{3 \pi} \rho_0 |u_{\text{1,JP}}|^3 A_s \left[(K_{\text{exp},s}+K_{\text{con},s}) + \left(\frac{A_s}{A_b}\right)^2 (K_{\text{con},b}+K_{\text{exp},b})  \right].
\label{eq:dE}
\end{align}

The optimal jet pump design is such that it will establish the required time-averaged pressure drop to cancel Gedeon streaming with a minimal amount of acoustic power dissipation. According to the quasi-steady approximation, this can be established by maximizing the difference between the contraction and expansion minor loss coefficients, while keeping the sum of the minor loss coefficients to a minimum. Designs such as the one presented in Fig. \ref{fig:jetpumpgeom} have been based on this analysis. However, recent work has shown that the accuracy of the quasi-steady approximation is limited and that factors other than the expansion and contraction minor losses play an important part. \cite{Oosterhuis2015,Petculescu2003,Smith2003a} Possibly, the most compelling example of this is the effect of the taper angle on the jet pump performance. Petculescu and Wilen have experimentally shown that a larger taper angle increases the minor losses for flow in the diverging jet pump direction. \cite{Petculescu2003} This will result in a significant performance drop due to the smaller time-averaged pressure drop and the higher time-averaged acoustic power dissipation. They suggest that the dependency on the taper angle is caused by a shift in the axial point, and therewith the local radius, at which flow separation occurs. However, they could not provide clear evidence of the existence of flow separation with their experimental set-up. 

In a recent numerical study, the time-averaged velocity fields inside several jet pump geometries have been examined, and it was concluded from the axial streaming field that flow separation had indeed occurred. \cite{Oosterhuis2015} The separated flow was only observed above certain displacement amplitudes and its onset was found at lower displacement amplitudes for larger taper angles. The study also showed that a severe decrease in jet pump performance started at approximately the same displacement amplitude as the onset of flow separation. It was therefore suggested that the latter is the cause of the problematic performance drop observed above certain displacement amplitudes. 

Next to axial streaming resulting from flow separation, it is necessary to examine the flow field in the vicinity of the separation in further detail. This will provide more insight into the influence of flow separation on the performance of jet pumps, which can subsequently be used to improve jet pump designs. In the following, a clear definition of separation in oscillatory flows is sought to serve as a basis for this investigation.

\subsection{Flow separation}
\label{sec:lit_flow_sep}

In an element with increasing cross-sectional area, such as the diverging direction of a jet pump, the boundary layer can separate due to the adverse pressure gradient overcoming the kinetic energy of the fluid. \cite{Petculescu2002} In steady flow, the separation is often characterized by vanishing wall shear stresses and flow reversal at the most upstream point. Downstream of this point the flow is separated, resulting in recirculation and a wake of disturbed flow that generates minor losses. However, Despard and Miller have experimentally shown that flow reversal in an oscillatory flow occurs periodically at all surface points in the adverse pressure gradient regime, thus also upstream of the wake formation. \cite{Despard1971} They stress that the steady flow characterization of flow separation is therefore unsuitable for oscillatory flow, and they propose to characterize flow separation with the surface point that first experiences flow reversal during the period of oscillation. This point can uniquely be determined and wake formation was found to invariably begin in the immediate neighborhood of this point. Simpson agreed that ``it is too narrow a view" to characterize flow separation with vanishing surface shear stresses and flow reversal in oscillatory flow, since flow reversal can occur without the separation of the boundary layer. \cite{Simpson1981} He emphasizes that separation must mean the complete breakdown or ``breakaway" of the boundary layer, which subsequently has a significant interaction with the bulk flow. It therefore seems necessary to extend the characterization of Despard and Miller with a measure that identifies the breakaway of the boundary layer. This could be done with the use of the time-averaged velocity fields inside the jet pump, as was shown in previous work. \cite{Oosterhuis2015}

King and Smith performed experiments with a purely oscillatory flow in a diffuser to investigate flow separation and minor losses for different operating conditions and diffuser angles. \cite{King2011} They state that the minor losses are mainly due to random motions generated after the flow separates, and that it is therefore reasonable to assume that a longer duration of flow separation over a larger spatial extent results in bigger minor losses. Due to the analogous adverse pressure gradient in a diffuser and a jet pump, their results show that one should aim to shorten the duration and spatial extent of flow separation in the diverging direction of a jet pump to increase its performance. King and Smith have shown that this can be accomplished by decreasing the taper angle of the geometry and by operating at either a higher Reynolds number for a given displacement amplitude or a lower displacement amplitude for a given Reynolds number. \cite{King2011} 

Schlichting had a more general view when he stated that the flow separation point is highly dependent on the adverse pressure gradient in the boundary layer, and that this pressure is imposed by the pressure of the bulk flow. \cite{Schlichting1968} This demonstrates that it is possible to control flow separation by geometrical variations that alter the adverse pressure gradient of the bulk flow. An example of this is the aforementioned decrease of the taper angle, but this might lead to undesirably large jet pumps. To increase the performance of jet pumps while keeping them compact, it therefore seems most effective to aim at reducing the adverse pressure gradient only in the immediate vicinity of the point of flow separation.

In this paper we investigate flow separation in the adverse pressure gradient regime of jet pumps using a computational fluid dynamics (CFD) model, which will be described in Sec. \ref{sec:modeling}. In Sec. \ref{sec:flow_sep_charac} we propose a measure to characterize flow separation in jet pumps and use this to examine flow separation and its effect on the jet pump performance. In Sec. \ref{sec:increased_trans_length}, jet pump designs that aim to suppress flow separation and increase the jet pump performance are investigated. Subsequently, partly due to the increased length of the proposed designs, the extent jet pumps can be made more compact without reducing their performance is investigated in Sec. \ref{sec:cut_cases}. Finally, an experimental validation of the CFD model for one of the proposed jet pump designs is given in Sec. \ref{sec:exp_validation}.    

\section{Modeling}
\label{sec:modeling}
In order to predict the performance of several jet pump designs, numerical simulations are performed. For this purpose, a two-dimensional axisymmetric CFD model similar to previous work\cite{Oosterhuis2015} is developed using the commercial software package ANSYS CFX 14.5. \cite{ANSYS2011} The computational domain with the applied boundary conditions and the analysis of the transient results are described here. An experimental validation of the model is presented in Sec. \ref{sec:exp_validation}.

\subsection{CFD model}
\label{sec:CFD_model}
All jet pump geometries are housed within an outer tube with radius $R_0 = $ \SI{30}{\mm} and length $L_0 = $ \SI{500}{\mm} on either side of the jet pump. The working fluid is air at a mean pressure of $p_0 = $ \SI{1}{atm} and a mean temperature of $T_0 = $ \SI{300}{\K}, which corresponds to the conditions in the experimental set-up (see Sec. \ref{sec:exp_validation}).

The unsteady, fully compressible Navier-Stokes equations are used to predict the transient flow field in the entire computational domain. To form a closed system of equations, the ideal gas law is used as an equation of state and the total energy equation including the viscous work term is used to describe the energy transport. \cite{ANSYS2012} No turbulence modeling is applied as the presented results all fall within the regime of laminar flow (see Sec. \ref{sec:laminar_flow}). The governing equations are discretized using a high resolution advection scheme in space and a second order backward Euler scheme in time. For each case, ten wave periods are simulated with 1000 time-steps per period. The experimental validation (Sec. \ref{sec:exp_validation}) is performed at an acoustic frequency of \SI{80}{\Hz}, while all other simulations are performed at an acoustic frequency of \SI{20}{\Hz}. For the latter, this results in a time-step of $\Delta t = 5 \cdot 10^{-5} $ \SI{}{\s}.  

\begin{figure}
	\centering
	\includegraphics[width=.5\textwidth]{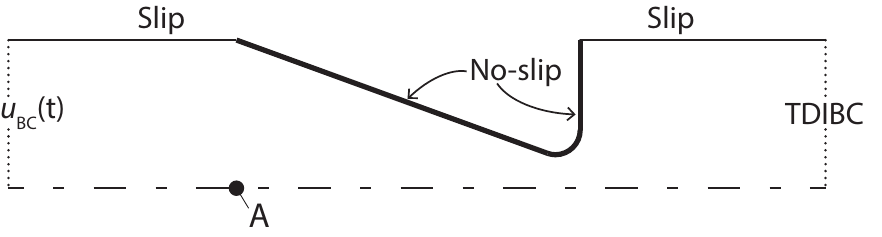}
	\caption{Schematic of the computational domain with the applied boundary conditions (not to scale). Point A represents the position where the time-averaged velocity $u_{2,L}$ is measured to detect flow separation (Sec. \ref{sec:charac_breakaway}). Dotted lines at the left and right hand side depict the local boundary conditions and the bottom dashed line represents the symmetry axis.}
	\label{fig:modeling_schematic}
\end{figure}

A schematic of the computational domain with the applied boundary conditions is given in Fig. \ref{fig:modeling_schematic}. At the left boundary of the domain, the acoustic wave is generated using a sinusoidal boundary condition with a specified velocity amplitude and frequency. To ensure a traveling wave, a time-domain impedance boundary condition (TDIBC) based on the work of Polifke et al. is applied at the right boundary of the domain. \cite{Huber2008a,Kaess2008} This approach sets the local pressure according to the specified complex reflection coefficient ($|R| = 0$ for a traveling wave). More details about the boundary condition and its implementation can be found in the work of Van der Poel, which has been carried out as part of this research. \cite{VanderPoel2013}  Note that the described numerical set-up ensures a zero mean mass flux across the boundaries of the computational domain.

Fig. \ref{fig:modeling_schematic} shows that an adiabatic no-slip boundary condition is applied at the jet pump wall. Because the viscous wall losses of the jet pump are the sole interest for this work, an adiabatic slip boundary condition is used at the outer tube wall. The axisymmetry is established by applying symmetry boundary conditions on the planes normal to the azimuthal direction. 

\subsection{Computational mesh}
The computational mesh is validated in previous work and details about the spatial discretization are only briefly repeated here. \cite{Oosterhuis2015} Within a region of \SI{50}{\mm} from the jet pump, the mesh is unstructured and consists of triangular and quadrilateral elements with a maximum size of \SI{1}{\mm}. A local refinement of \SI{0.5}{\mm} is used near the jet pump waist and at the transition from the no-slip to the slip boundary condition at the left hand side of the jet pump. Furthermore, an inflation layer is applied such that a minimum of 15 elements is present within a distance of $2 \pi \delta_v$ from the jet pump wall. Here $\delta_v$ is the viscous penetration depth given by: $\delta_v = \sqrt{\mu_0/\pi \rho_0 f}$. In the remaining part of the domain, a structured mesh with only quadrilateral elements is constructed. In the radial direction the mesh size is fixed at \SI{1}{\mm}, whereas in the axial direction the mesh size increases from \SI{1}{\mm} near the jet pump section up to \SI{10}{\mm} at the boundaries of the domain. The typical mesh size is \SI{40000}{} nodes for the geometries presented in this work. This results in a single core computational time of approximately \SI{40}{\hour} on an Intel Core i7 CPU.   

\subsection{Data analysis}
\label{sec:data_analysis}
To investigate the jet pump performance, the transient solution fields of the last five wave periods are processed. A discrete Fourier transform at the specified wave frequency is used to acquire the complex amplitudes of the flow field variables. For the sinusoidal jet pump velocity, the displacement amplitude $\xi_1$ inside the jet pump waist is calculated by
\begin{equation}
\xi_1 = \frac{|u_{\text{1,JP}}|}{2 \pi f},
\label{eq:disp_amp}
\end{equation} 
where $|u_{\text{1,JP}}|$ is the velocity amplitude in the jet pump waist determined by an area-weighted average over the cross section. Similar to the dimensionless stroke length $L_0/h$ of Smith and Swift, \cite{Smith2003a} the displacement amplitude in the jet pump waist is scaled with the local diameter ($D_s = 2R_s$) to acquire the Keulegan-Carpenter number,
\begin{equation}
KC_D = \frac{\xi_1}{D_s}.
\label{eq:KCd}
\end{equation} 

The time-averaged pressure drop $\Delta p_2$ over the jet pump is calculated by taking the difference of the spatially averaged, time mean pressure on either side of the jet pump. The time-averaged acoustic power dissipation $\Delta \dot{E}_2$ is determined similarly from the difference in acoustic power, which is given by \cite{Swift2002a} 
\begin{equation}
\dot{E}_2(x) = \frac{1}{2}\Re[\widetilde{p_1}(x)U_1(x)],
\end{equation}
where $\widetilde{p_1}(x)$ is the complex conjugate of the area-averaged pressure amplitude and $U_1(x)$ is the complex volume flow rate. To be able to compare cases with different geometries and operating conditions, the performance indicators $\Delta p_2$ and $\Delta \dot{E}_2$ are scaled according to the work of Smith and Swift, \cite{Smith2003a}
\begin{align}
\Delta p_2^* &= \frac{8 \Delta p_{2}}{\rho_0 |u_{\text{1,JP}}|^2} 
\label{eq:dp2star}, \\
\Delta \dot{E}^*_2 &= \frac{3 \pi \Delta \dot{E}_{2}}{\rho_0 A_s |u_{\text{1,JP}}|^3}.
\label{eq:dim_power_dissipation}
\end{align}
An effective jet pump should generate the required amount of dimensionless pressure drop to cancel Gedeon streaming, while the dimensionless acoustic power dissipation is minimal. Hence, the ratio of $\Delta p_2^*$ over $\Delta \dot{E}^*_2$ is referred to as the effectiveness of the jet pump.

\subsection{Laminar flow}
\label{sec:laminar_flow}
The modeling of turbulence continues to be a major issue in predicting flow separation. \cite{Cherrye2008} Turbulence models, such as the wide variety of low-Reynolds number k-epsilon models, fail to reproduce the near-wall flow characteristics in detail. \cite{Michelassi1993} Therefore, it is ensured in this work that the flow remains laminar within the entire computational domain so that no turbulence modeling is needed. This is done through the use of the acoustic Reynolds number, which is defined as
\begin{equation}
Re_\delta = \frac{|u_1| \rho_0 \delta_v }{\mu_0}.
\end{equation} 
The maximum acoustic Reynolds number occurs at the jet pump waist, where the velocity amplitude is the largest. The critical Reynolds number for laminar oscillatory pipe flow is derived by Ohmi and Iguchi as \cite{Ohmi1982}
\begin{equation}
Re_c = 305\left(\frac{D}{\delta_v}\right)^{1/7}.
\end{equation}
For all results presented in this work, the acoustic Reynolds number in the jet pump waist is kept smaller than the critical Reynolds number. This verifies that the flow remains within the laminar regime, and thus no additional turbulence modeling is needed.

\section{Characterization of flow separation}
\label{sec:flow_sep_charac}
To characterize and investigate the separation of flow in the diverging direction of jet pumps, a reference geometry based on the design of Fig. \ref{fig:jetpumpgeom} is used. Table \ref{tab:reference_geometry} lists the geometrical parameters of the reference geometry. Note that the big opening is eliminated, i.e. $R_b = R_0$, such that the tapered surface of the jet pump reaches the outer tube (see Fig. \ref{fig:modeling_schematic}). This is done to independently investigate the separation of flow and the extent to which jet pumps can be made more compact (Sec. \ref{sec:cut_cases}). This also negates the effect of vortex generation at the big opening from the flow separation analysis. 

\begin{table}
	\centering
		\caption{Dimensions of the simulated reference geometry. The corresponding jet pump design is depicted in Fig. \ref{fig:jetpumpgeom}.}
		\label{tab:reference_geometry}
		\begin{tabular}{ll}
			$L_\text{JP}$ & \SI{91.5}{\mm}  \\
			$R_0$ & \SI{30}{\mm} \\
			$R_b$ & \SI{30}{\mm}  \\
			$R_s$ & \SI{7}{\mm} \\
			$R_c$ & \SI{5}{\mm} \\
			$\alpha$ & 15$^\circ$ \\
		\end{tabular}
\end{table}

First, a measure to determine the breakaway of the boundary layer is specified using the numerical results of the reference geometry. This is followed with an analysis of the time and axial point at which flow reversal first occurs along the jet pump wall. Both results are then used to investigate the influence of flow separation on the jet pump performance.       

\subsection{Boundary layer breakaway}
\label{sec:charac_breakaway}
\begin{figure}
	\centering
	\includegraphics[width=.6\textwidth]{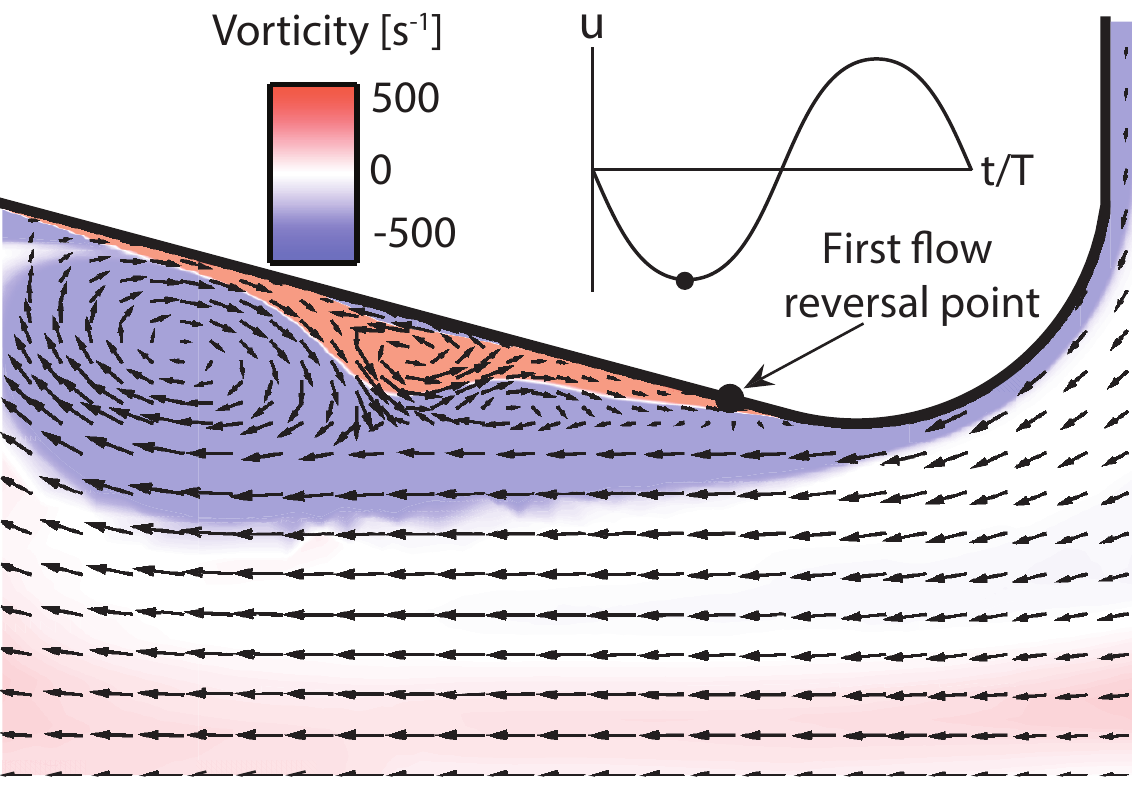}
	\caption{Vector plot of the velocity field for a section of the reference geometry where $t/T = 0.25$ and $KC_D = 3.5$. The underlaying shades depict the vorticity field.}
	\label{fig:vector_ref1}
\end{figure}

To characterize flow separation by the breakaway of the boundary layer, we focus on the half-period in which the bulk flow in the jet pump waist is moving leftward ($0 < t/T < 0.5$). Here, a period of time $T$ is defined to start when the bulk flow velocity $u$ at the jet pump waist is zero, and the flow starts moving leftward. To identify the boundary layer breakaway, we first look at the flow in the vicinity of the jet pump wall. It is observed in all investigated cases that the flow in the boundary layer changes direction, while the bulk flow is still leftward. Thus, the boundary layer is leading the bulk flow in phase. An example where the reversed flow has already occurred is given in Fig. \ref{fig:vector_ref1}. Here, the velocity field at the end of the leftward acceleration ($t/T = 0.25$) is shown for $KC_D = 3.5$. At this Keulegan-Carpenter number, the first flow reversal occurred at $t/T \approx 0.08$. This initially reversed flow has been observed to move rightward along the wall until it reached the contracting bulk flow at the jet pump waist. At this point, the reversed flow was prohibited from moving any further and it was forced to move downwards and recirculate into the bulk flow. This recirculation continues up to $t/T = 0.5$, and induces a stream of vortices that propagate leftward with the bulk flow. These vortices and the resulting wake formation can be seen at the instant in time depicted in Fig. \ref{fig:vector_ref1}, and are found to be present for all cases investigated in this work. For example, a transient representation of the wake formation process at $KC_D=7.3$ is available in Mm.~1.

\noindent\hspace{10pt}\begin{minipage}{\linewidth}
\vspace{1cm}
\flushleft\textit{Mm.~1: Evolution of the flow field in the full reference jet pump geometry at $KC_D=7.3$ for $0 < t/T < 0.7$. This is a file of type ``mpg'' (6.0~Mb) and can be found with the \doi{ 10.1121/1.4939490}.}
\vspace{1cm}
\end{minipage}

Two distinct cases of wake formation as a function of displacement amplitude can be identified. For $KC_D \le 1.7$, the induced vortices propagate only slightly towards the left, after which the bulk flow changes direction. The vorticity remains present for some time, but diminishes during the half-period of rightward bulk flow. This only results in a deviation from a purely oscillatory flow in the direct vicinity of the jet pump waist. In the rest of the flow field, no significant disturbances are caused by this vorticity. For $KC_D > 1.7$, the vorticity does not diminish completely during the rightward bulk flow. This causes time-averaged streaming inside the jet pump, significantly disturbing the purely sinusoidal flow. The latter case is characterized as flow separation due to the large disturbance of the bulk flow caused by the breakaway of the boundary layer. It can be identified by a negative time-averaged velocity at the symmetry axis and a positive time-averaged velocity inside the rest of the jet pump, as shown in previous work \cite{Oosterhuis2015}. The transition between the two cases occurs when the vorticity induced by the flow reversal remains present during the rightward acceleration of the bulk flow ($0.5 < t/T < 0.75$). If the vorticity is still present after $t/T = 0.75$, the flow at the symmetry axis is found to remain predominantly negative and flow separation has occurred. As a measure for the latter, we state that flow separation has occurred if
\begin{equation}
\frac{u_{2,L}}{|u_{1,FF}|} \le -1,
\label{eq:flow_sep}
\end{equation}
where $u_{2,L}$ is the time-averaged axial velocity at the point just left of the big jet pump opening, as depicted by point A in Fig. \ref{fig:modeling_schematic}. $|u_{1,FF}|$ is the amplitude of the far field oscillatory flow at the left hand side of the jet pump. For undisturbed oscillatory flow, this velocity amplitude is approximately equal in the entire outer tube to the left of the jet pump. Eq. \ref{eq:flow_sep} therefore states that flow separation has occurred if there is a negative streaming velocity at the symmetry axis that is bigger than the local amplitude of the oscillatory flow. It is found that the ratio of Eq. \ref{eq:flow_sep} is slightly positive if no flow separation occurs and is significantly less than $-1$ if flow separation does occur. It therefore seems to be an appropriate measure to identify the breakaway of the boundary layer and the resulting flow separation.  

\subsection{First flow reversal}
\label{sec:flow_reversal}
So far, it has been shown that flow reversal along the jet pump wall occurs during leftward bulk flow. Despard and Miller state that the time and axial point at which this flow reversal first occurs is important to characterize flow separation. \cite{Despard1971} Therefore, we have written an algorithm that searches for the point along the jet pump wall where the flow changes sign during negative bulk flow ($0 < t/T < 0.5$). The axial point and time at which this occurs are averaged over the last five wave periods.

It is found that the axial point of first flow reversal is situated \SI{1.9} {\mm} left of the jet pump waist, where the local radius is \SI{7.4} {\mm} (see Fig. \ref{fig:vector_ref1}). This result is found to be independent of the displacement amplitude and oscillation frequency, which was also reported by Despard and Miller in their experimental work. \cite{Despard1971} The leftward wake formation starts in the immediate vicinity of the first flow reversal point, although this is hard to clarify for the reference geometry as this point is close to the jet pump waist. This will become clearer in Sec. \ref{sec:increased_trans_length}, where geometries are presented that shift the first flow reversal point away from the jet pump waist. 

\begin{figure}
	\centering
	\includegraphics[width=.5\textwidth]{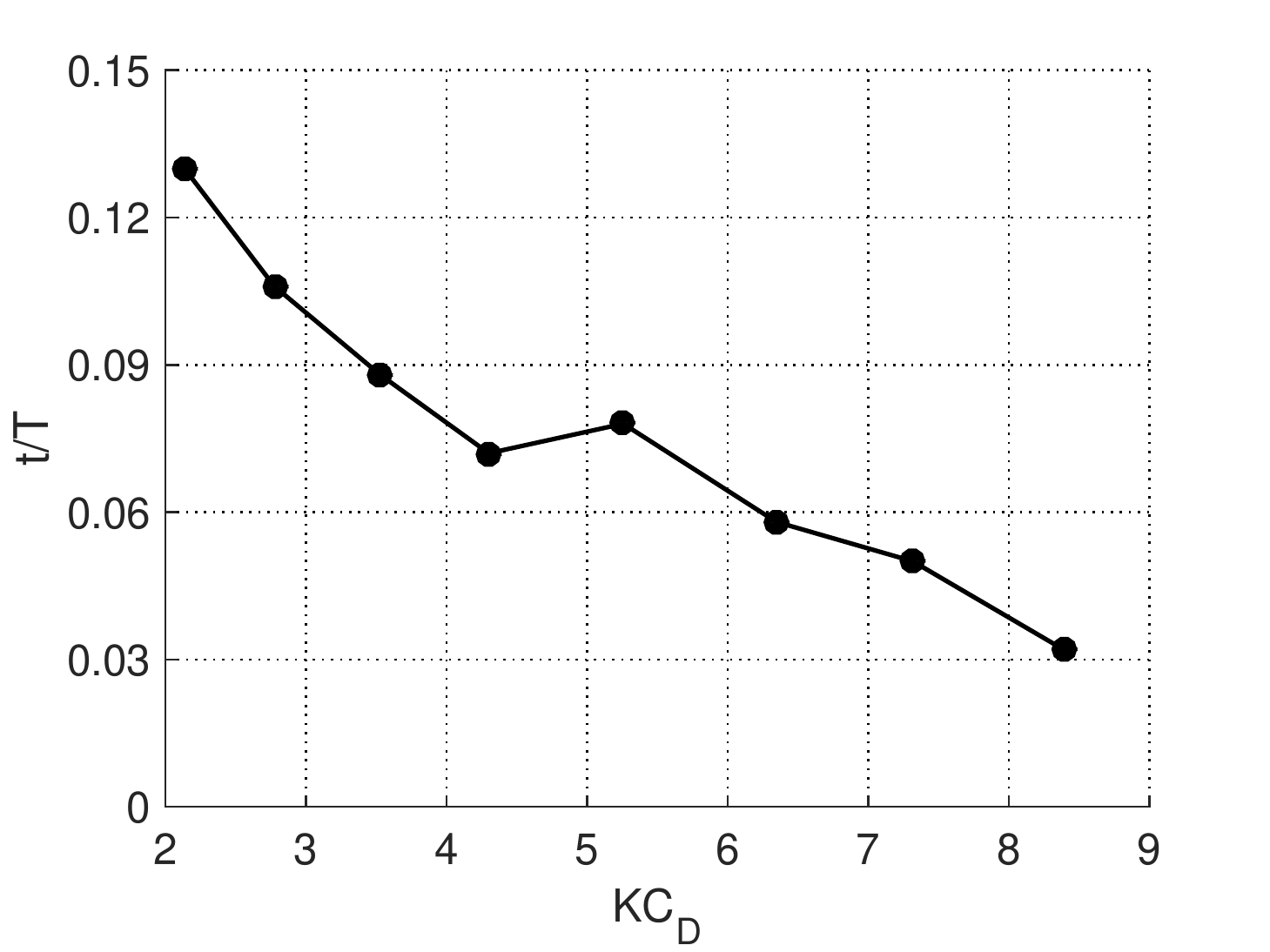}
	\caption{Time at which flow reversal was first found as a function of $KC_D$. Only the cases with flow separation for the reference geometry are depicted.}
	\label{fig:ref_flow_sep_time}
\end{figure}

The time at which flow reversal first occurs is depicted in Fig. \ref{fig:ref_flow_sep_time} as a function of $KC_D$ for the cases with flow separation. For increasing displacement amplitude, the first occurrence of flow reversal shifts to earlier on in the period. Following King and Smith, it is therefore reasonable to state that the longer duration of flow separation for increasing displacement amplitudes generates more minor losses in the leftward direction. \cite{King2011} In turn, this will reduce the time-averaged pressure drop across the jet pump when compared with the situation where no flow separation would occur. To investigate the impact of reducing the time of first flow reversal on the jet pump performance, the dimensionless pressure drop in the flow separation regime is examined in the following section.  

\subsection{Effect of flow separation on performance}
\label{sec:charac_flow_sep_performance}
\begin{figure}
	\centering
	\includegraphics[width=.55\textwidth]{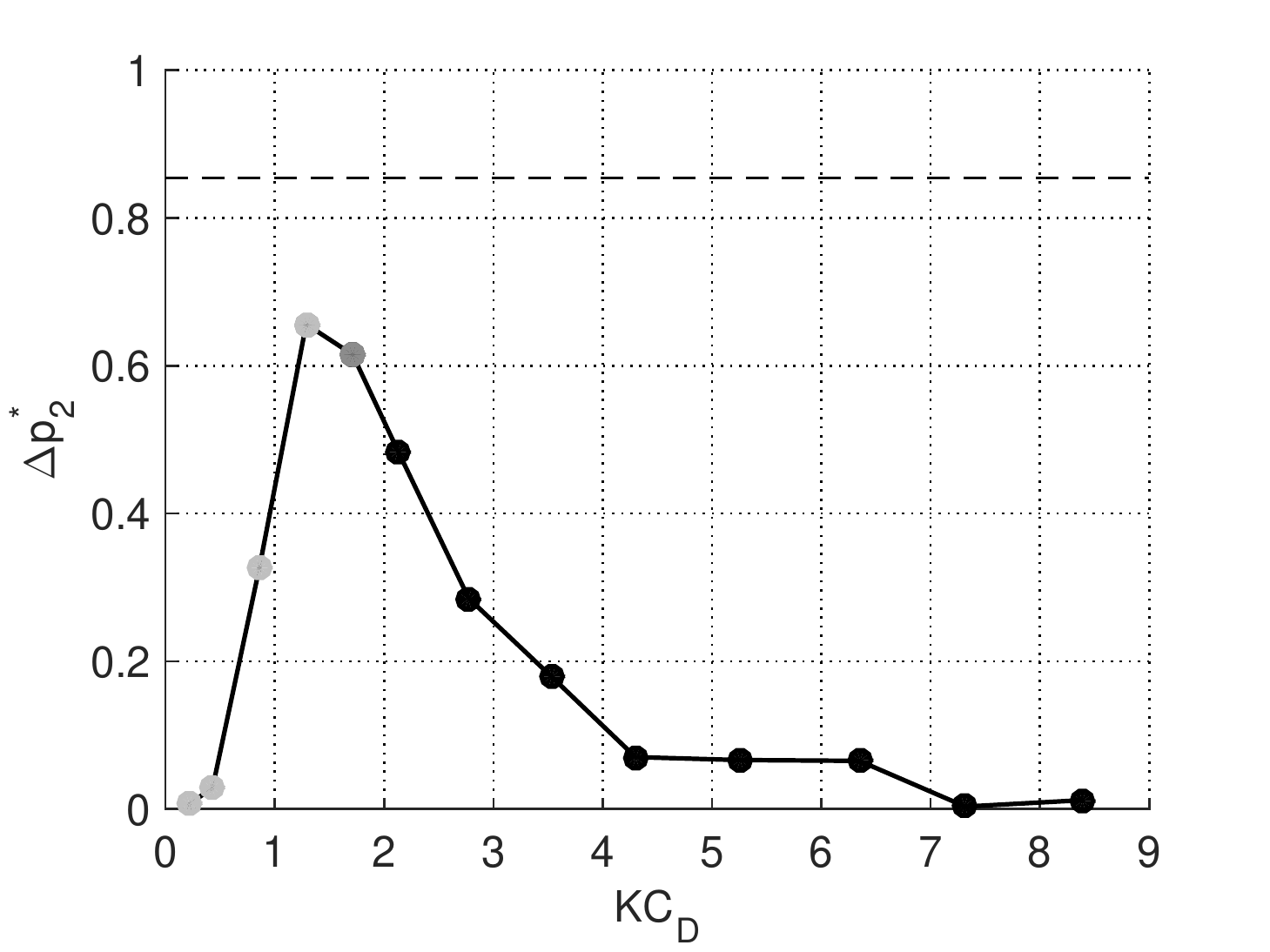}
	\caption{Dimensionless pressure drop as a function of $KC_D$ for the reference geometry. Three flow regimes are shown; no flow separation (light gray markers), partial flow separation (dark gray marker), and full flow separation (black markers). The dashed black line represents the quasi-steady approximation.}
	\label{fig:dp_ref}
\end{figure}

Now that flow separation has been characterized in terms of the breakaway of the boundary layer and flow reversal, its influence on the jet pump performance can be investigated. The dimensionless pressure drop as a function of $KC_D$ for the reference geometry is depicted in Fig. \ref{fig:dp_ref}. For $KC_D < 0.5$, there is no significant dimensionless pressure drop. Thereafter it starts increasing rapidly for larger displacement amplitudes, up to a maximum at $KC_D = 1.3$. After this maximum, a severe decay in the dimensionless pressure drop for increasing displacement amplitudes occurs. Therefore, there is only a small range of displacement amplitudes in which the jet pump effectively produces the required pressure drop to cancel Gedeon streaming. Moreover, even if the jet pump is operated around the point of maximum performance, there is a large sensitivity of the performance on small changes in the operating conditions. To suppress Gedeon streaming, it can therefore be difficult to induce the correct amount of time-averaged pressure drop with this jet pump design. In the search of more robust jet pump designs, that do not have a small range of effective performance, it is important to understand why the dimensionless pressure drop reduces so severely above certain displacement amplitudes.

The black markers in Fig. \ref{fig:dp_ref} depict the dimensionless pressure drop of the cases where flow separation according to Eq. \ref{eq:flow_sep} was found. All cases are situated in the declining part of the dimensionless pressure drop, i.e. after the maximum at $KC_D = 1.3$. This shows that there is a direct relation between flow separation and a decrease in jet pump performance. For increasing displacement amplitude, the negative effect of flow separation on the dimensionless pressure drop is increasing. As described in the previous section, this can be attributed to flow reversal occurring earlier on in the period (see Fig. \ref{fig:ref_flow_sep_time}), which induces more minor losses that reduce the dimensionless pressure drop.

The light gray markers in Fig. \ref{fig:dp_ref} show the cases were no flow separation is present and the dimensionless pressure drop rises with increasing $KC_D$. The dimensionless pressure drop does not reach the value of the quasi-steady approximation (dashed black line). This can be attributed to flow separation, recalling that only expansion and contraction minor losses at the jet pump openings are included in the quasi-steady approximation. The onset of flow separation occurs at $KC_D = 2.1$, which is such a low Keulegan-Carpenter number that the dimensionless pressure drop can not increase up to the value that it could without flow separation. To acquire a higher maximum dimensionless pressure drop, the onset of flow separation should therefore be shifted to a higher displacement amplitude, as attempted in Sec. \ref{sec:increased_trans_length}.  

\begin{figure}
	\centering
	\includegraphics[width=.6\textwidth]{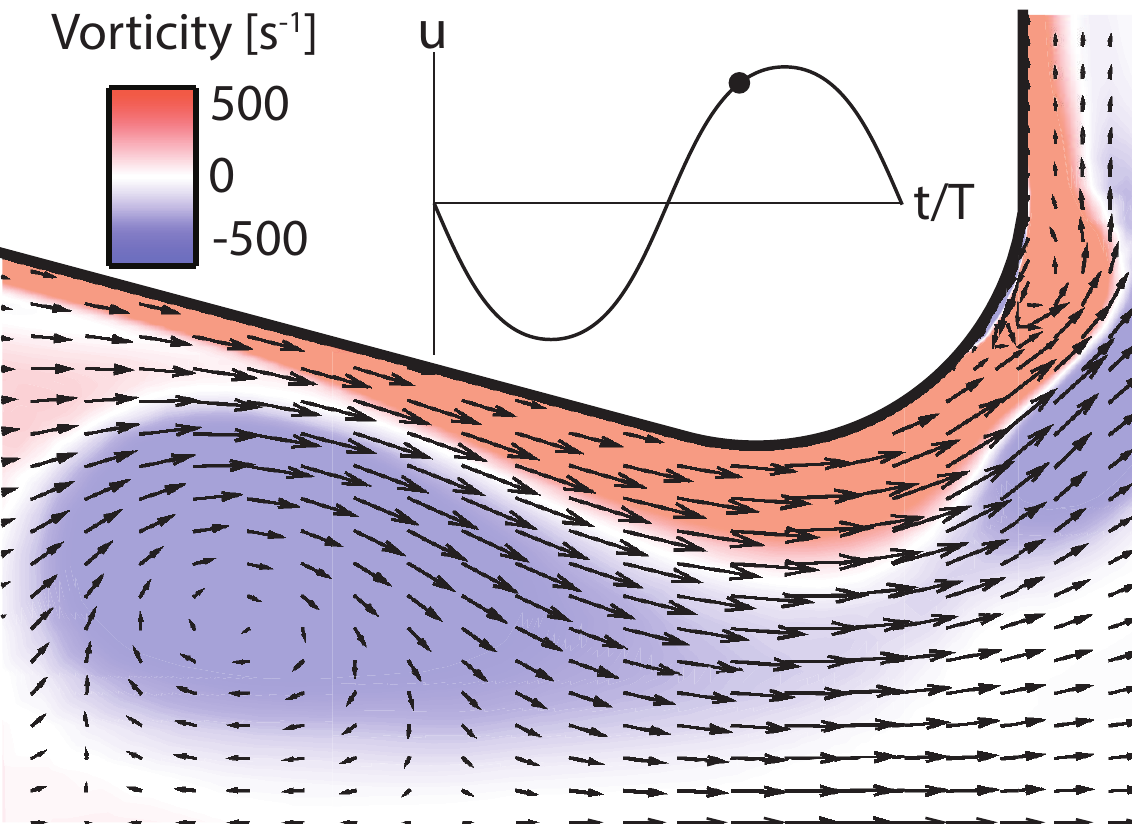}
	\caption{Vector plot of the velocity field for a section of the reference geometry where $t/T = 0.65$ and $KC_D = 1.7$. The underlaying shade depict the vorticity field.}
	\label{fig:vector_ref2}
\end{figure}

So far, the decreasing dimensionless pressure drop has been attributed to flow separation, but why the dimensionless pressure drop decreases before the onset of flow separation according to Eq. \ref{eq:flow_sep} has not been explained. First of all, it should be clear that a decay of the dimensionless pressure drop does not represent a decay of the time-averaged pressure drop; it means that the latter increases in a manner less than quadratic. We therefore seek to explain why the time-averaged pressure drop changes from increasing more than quadratically (light gray markers in Fig. \ref{fig:dp_ref}) to less than quadratically (dark gray marker in Fig. \ref{fig:dp_ref}). The velocity field of the first case that experiences a decrease in the dimensionless pressure drop is shown during the rightward bulk flow in Fig. \ref{fig:vector_ref2}. At the depicted instant in time ($t/T = 0.65$), the generation of the rightward expansion minor losses has just started. This can be seen by the flow reversal and vorticity along the right hand side of the jet pump wall. Additionally, the vorticity induced during the leftward bulk flow is still present just to the left of the jet pump waist. This vorticity will diminish before $t/T = 0.75$, and will not cause time-averaged streaming inside the jet pump. However, it does prevail up to the time where the generation of the expansion minor losses in the rightward direction starts. Therefore, the vorticity reduces the strength of the rightward bulk flow during this period, and therewith the minor losses in the rightward direction that are predominantly responsible for the time-averaged pressure drop across the jet pump. In contrast, for the cases with an increasing dimensionless pressure drop ($KC_D \le 1.3$), the vorticity is completely diminished during the time at which the expansion minor losses start. This shows that the interaction between the two causes a decrease in the dimensionless pressure drop for displacement amplitudes even lower than those where flow separation is present. This situation is referred to as partial flow separation, since there is a short interaction of the induced vorticity with the bulk flow, but no time-averaged streaming inside the jet pump occurs.

The negative effect on the time-averaged pressure drop for the partial flow separation is also present for the case of full flow separation. Due to the non-diminishing vorticity and resulting negative streaming at the symmetry axis, the effect will even be larger. Therefore, next to the wake formation that increases the minor losses in the leftward direction, flow separation also decreases the minor losses in the rightward direction. This shows another incentive to reduce flow separation in an effort to increase the performance of jet pumps.

\section{Reduction of flow separation}
\label{sec:reduction_flow_sep}
In the previous section it is shown that the point of first flow reversal is close to the jet pump waist for the reference geometry. In an effort to reduce the negative effects of flow separation on the jet pump performance, we propose design alterations that reduce the adverse pressure gradient in the vicinity of this point.

\subsection{Increased transition length}
\label{sec:increased_trans_length}
\begin{figure}
	\centering
	\includegraphics[width=.6\textwidth]{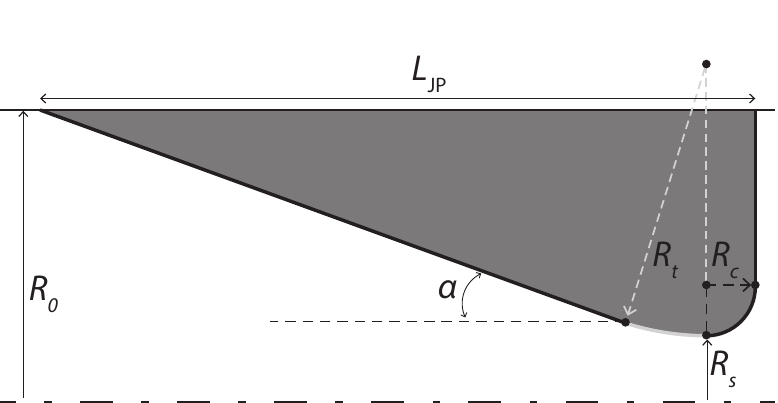}
	\caption{Axisymmetric representation of the adjusted jet pump with parameters that define the geometry. Bottom dashed line indicates center line, top solid line indicates outer tube wall. $R_t$ is varied to acquire different geometries.}
	\label{fig:geom_rc2}
\end{figure}

A reduced adverse pressure gradient near the point of first flow reversal is ensured by creating a longer and smoother transition from the jet pump waist to the tapered surface. For the reference geometry, this transition is done by a circle trajectory with the same radius as the radius of curvature of the small opening. For the adjusted designs, the transition radius $R_t$ is increased while keeping the radius of curvature $R_c$ of the small jet pump opening constant. An example of such a design is depicted in Fig. \ref{fig:geom_rc2}, where $R_t$ = \SI{28} {\mm}, $L_\text{JP}$ = \SI{94.5} {\mm}, and the other parameters are the same as the reference geometry. When compared with the reference geometry ($R_t$ = \SI{5} {\mm} and $L_{\text{JP}}$ = \SI{91.5} {\mm}), this design has an increased length $\Delta L_{\text{JP}}$ of \SI{3} {\mm} due to the elongated transition length. By further varying $R_t$, different geometries are designed that are characterized by the extra transition length $\Delta L_{\text{JP}}$ they introduce with respect to the reference case. Six different geometries are examined: $\Delta L_{\text{JP}}$ = \SI{3}{\mm}, \SI{5} {\mm}, \SI{10} {\mm}, \SI{20} {\mm}, \SI{30} {\mm}, and \SI{50} {\mm}.    

\begin{figure}
	\centering
	\includegraphics[width=.6\textwidth]{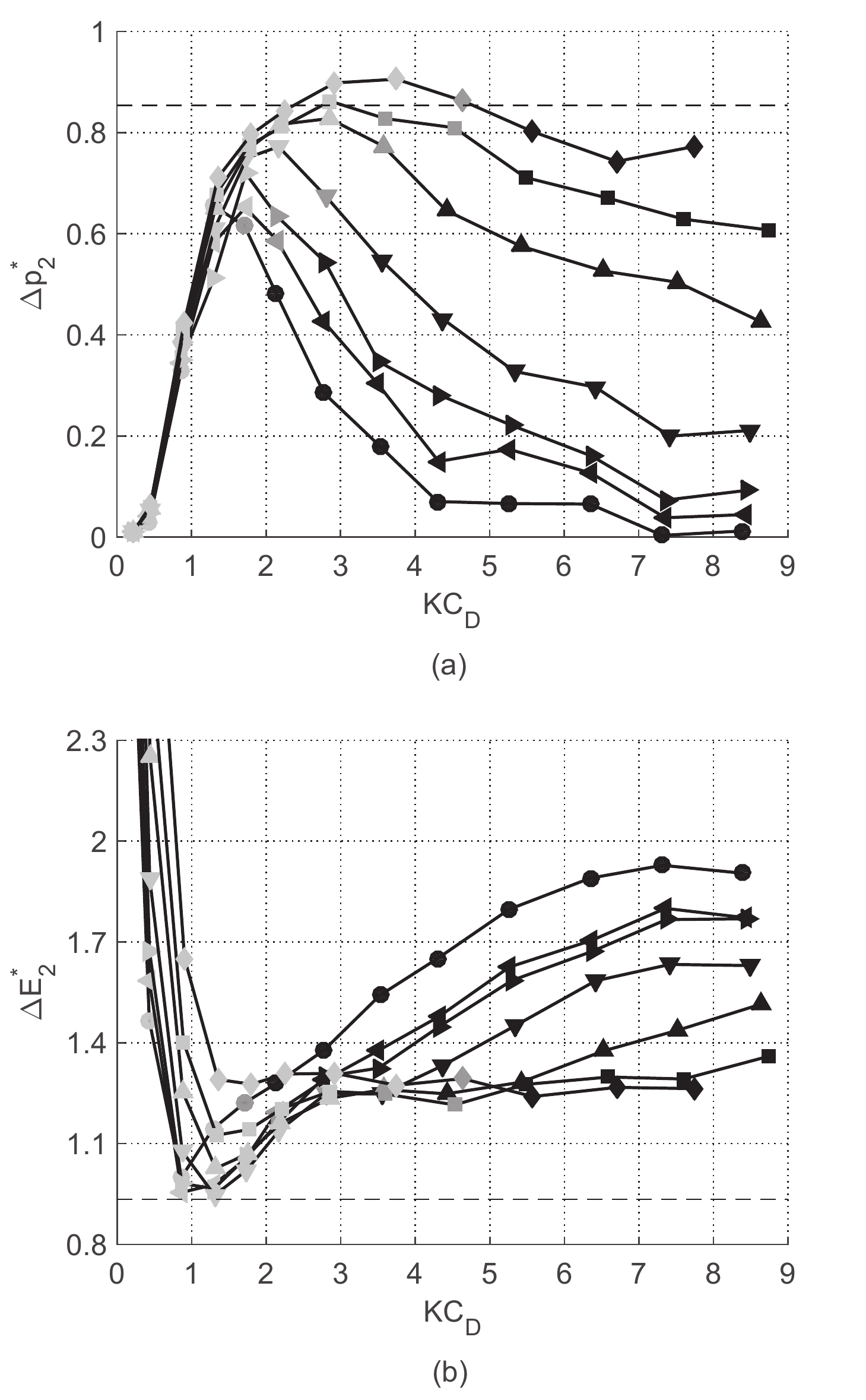}
	\caption{Dimensionless pressure drop (a) and acoustic power dissipation (b) as a function of $KC_D$ for the transition lengths: $\Delta L_{\text{JP}}$ = \SI{0} {\mm} ({\Large $ \bullet $}, reference), $\Delta L_{\text{JP}}$ = \SI{3} {\mm} ($\blacktriangleleft$), $\Delta L_{\text{JP}}$ = \SI{5} {\mm} ($\blacktriangleright$), $\Delta L_{\text{JP}}$ = \SI{10} {\mm} ($\blacktriangledown$), $\Delta L_{\text{JP}}$ = \SI{20} {\mm} ($\blacktriangle$), $\Delta L_{\text{JP}}$ = \SI{30} {\mm} ({\footnotesize $\blacksquare$}), and $\Delta L_{\text{JP}}$ = \SI{50} {\mm} ($\blacklozenge$). Three flow regimes are depicted: no flow separation (light gray markers), partial flow separation (dark gray markers), and full flow separation (black markers). The dashed black line represents the quasi-steady approximation.}
	\label{fig:transition_length_dp_de}
\end{figure}

The performance of the adjusted jet pump geometries and the reference geometry is depicted in Fig. \ref{fig:transition_length_dp_de} in terms of the dimensionless pressure drop (a) and acoustic power dissipation (b). All geometries show a very similar increase of the dimensionless pressure drop for increasing displacement amplitude up to $KC_D = 1.3$. At this Keulegan-Carpenter number, the displacement amplitude up to which no flow separation is present (light gray markers) is reached for the reference geometry. Here a maximum dimensionless pressure drop of $\Delta p^*_2 = 0.66$ is acquired. For the cases with an increased transition length, the region of no flow separation (light gray markers) is extended to a larger displacement amplitude compared with the reference geometry. Therefore, the dimensionless pressure drop continues to increase, resulting in an improvement of the maximum dimensionless pressure drop up to $\Delta p^*_2 = 0.91$ for $\Delta L_{\text{JP}}$ = \SI{50} {\mm}. It is worth noting that for the geometries with the largest transition lengths, the flow separation is postponed to such a displacement amplitude that the dimensionless pressure drop even reaches the value predicted by the quasi-steady approximation (dashed black line). Eventually, just as for the reference geometry, the dimensionless pressure drop decreases again for all geometries with increasing displacement amplitude. However, for a larger $\Delta L_{\text{JP}}$ the decrease of the dimensionless pressure drop is significantly reduced compared with a smaller transition length. This results in a broader range of displacement amplitudes where the dimensionless pressure drop is near its maximum for increasing $\Delta L_{\text{JP}}$. Therefore, a larger transition length shows a positive influence on the maximum dimensionless pressure drop as well as the robustness of the jet pump. 

In Fig. \ref{fig:transition_length_dp_de}(a), it can be seen that the initial increase of the dimensionless pressure drop up to a maximum occurs for all geometries in the regime where no flow separation is present (light gray markers). The regime of partial flow separation (dark gray markers) is first found at displacement amplitudes just above the maximum dimensionless pressure drop. This is followed by full flow separation (black markers) and a decrease of the dimensionless pressure drop. These results show that for increased transition lengths, the different flow regimes occur at the same characteristic points of the dimensionless pressure drop curve as for the reference geometry. This confirms the proposed characterization of flow separation and its influence on the jet pump performance. 

So far, we have only examined the dimensionless pressure drop, but the dimensionless acoustic power dissipation is also an important measure to quantify the jet pump performance. An effective jet pump will have a high pressure drop at the cost of a low acoustic power dissipation. In Fig. \ref{fig:transition_length_dp_de}(b) it can be seen that all geometries have an asymptotically high $\Delta \dot{E}^*_2$ for displacement amplitudes that approach zero. This is caused by the scaling of the acoustic power dissipation with the velocity amplitude cubed (see Eq. \ref{eq:dim_power_dissipation}), which is very small for these displacement amplitudes. The dimensionless acoustic power dissipation then drops quickly with increasing displacement amplitude. In the regime without flow separation (light gray markers), the cases with a larger transition length have more acoustic power dissipation. This can be explained by the increased jet pump length in the vicinity of the jet pump waist, where the velocity amplitude is the largest and therewith the most viscous losses occur. For increasing $KC_D$, flow separation (black markers) sets in causing the dimensionless acoustic power dissipation to increase significantly, especially for the cases with the smallest $\Delta L_{\text{JP}}$. As a result of this, a larger transition length is favorable for high displacement amplitudes in the flow separation regime, due to the reduced acoustic power dissipation. This can be explained by the reduction of the minor losses generated by flow separation compared with a case that has a smaller transition length. This same reason also ensured a larger dimensionless pressure drop for a higher $\Delta L_{\text{JP}}$ in the flow separation regime (black markers in Fig. \ref{fig:transition_length_dp_de}(a)).  

\begin{figure}
	\centering
	\includegraphics[width=.5\textwidth]{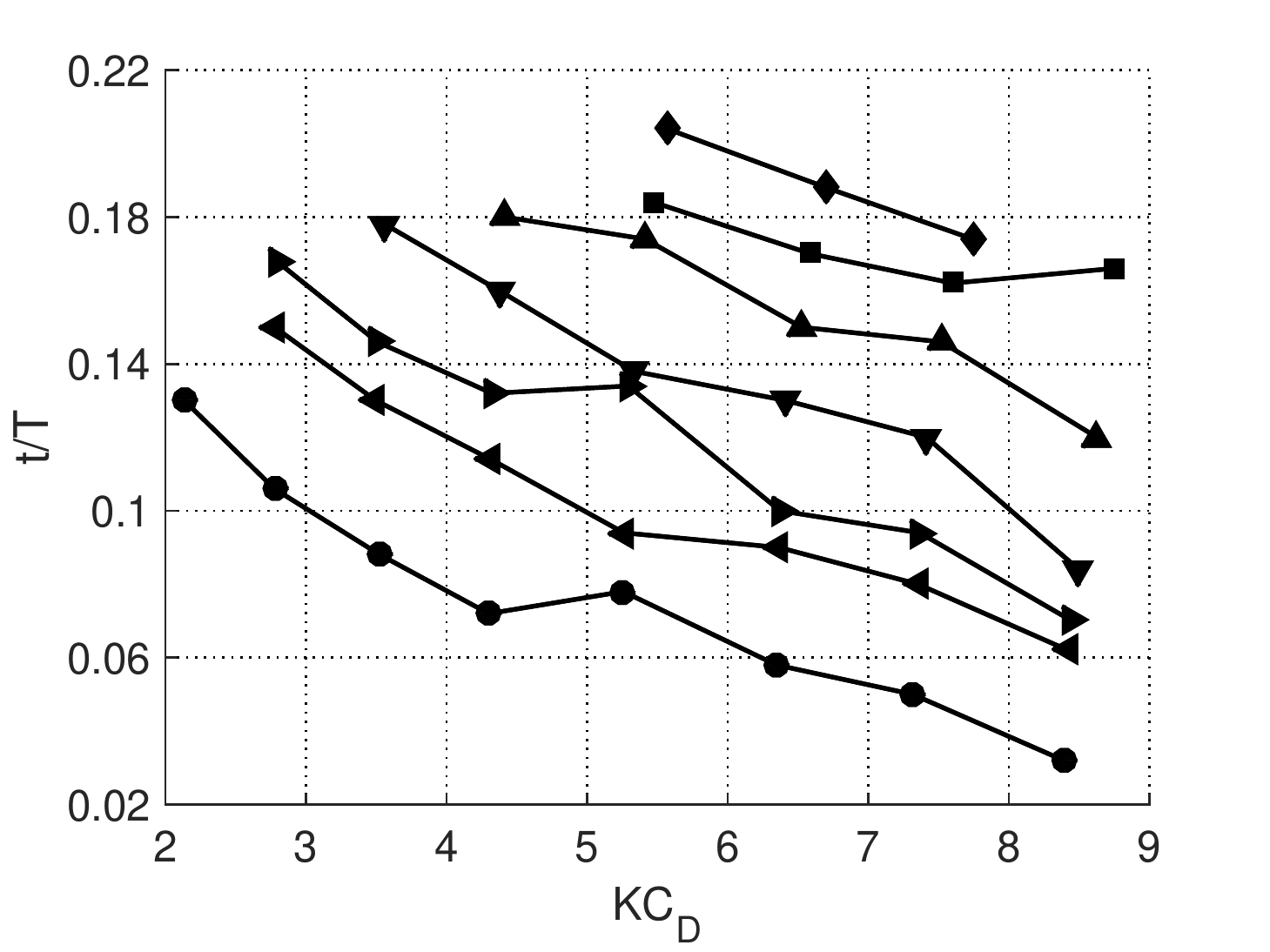}
	\caption{Time at which flow reversal was first found as a function of $KC_D$. Only the cases with flow separation are depicted. Symbols indicate the transition lengths:  $\Delta L_{\text{JP}}$ = \SI{0} {\mm} ({\Large $ \bullet $}, reference), $\Delta L_{\text{JP}}$ = \SI{3} {\mm} ($\blacktriangleleft$), $\Delta L_{\text{JP}}$ = \SI{5} {\mm} ($\blacktriangleright$), $\Delta L_{\text{JP}}$ = \SI{10} {\mm} ($\blacktriangledown$), $\Delta L_{\text{JP}}$ = \SI{20} {\mm} ($\blacktriangle$), $\Delta L_{\text{JP}}$ = \SI{30} {\mm} ({\footnotesize $\blacksquare$}), and $\Delta L_{\text{JP}}$ = \SI{50} {\mm} ($\blacklozenge$).}
	\label{fig:transition_length_flow_rev_time}
\end{figure}

To investigate why the minor losses induced by flow separation are reduced for increased transition lengths, the first point of flow reversal is identified for the cases where flow separation occurs. This process is visually depicted in Mm.~2, where a one--to--one comparison is shown of the flow reversal between the reference geometry (also shown in Mm.~1) and the case with $\Delta L_{\text{JP}}=\SI{20}{\mm}$ at $KC_D\approx 7.4$. Using the algorithm described in Sec.~\ref{sec:flow_reversal}, the time and point of first flow reversal are determined. For all cases, the time at which the first flow reversal is found is depicted in Fig. \ref{fig:transition_length_flow_rev_time} as a function of $KC_D$. All geometries show a decrease in the time of first flow reversal with increasing displacement amplitude, which will cause more minor losses in the leftward direction. This concurs with the reduction of the dimensionless pressure drop and the increase of the dimensionless acoustic power dissipation for an increasing displacement amplitude. 
For all displacement amplitudes in the flow separation regime, an increase in $\Delta L_{\text{JP}}$ leads to a delay in the time of first flow reversal. Therefore, by increasing the transition length, the minor losses generated by the wake formation of flow separation are reduced. This explains why the dimensionless pressure drop is higher and the dimensionless acoustic power dissipation is lower for larger transition lengths in the flow separation regime.  Next to that, it can be seen that the onset of flow separation occurs at a higher displacement amplitude for a larger transition length. This ensures that the maximum dimensionless pressure drop is increased and shifted to higher values of $KC_D$, as can be seen in Fig. \ref{fig:transition_length_dp_de}(a).

\noindent\hspace{10pt}\begin{minipage}{\linewidth}
\vspace{1cm}
\flushleft\textit{Mm.~2: Comparison of the flow fields in the reference jet pump geometry (at $KC_D=7.3$) and in the geometry with $\Delta L_{\text{JP}}=\SI{20}{\mm}$ (at $KC_D=7.5$) for $0 < t/T < 0.5$. This is a file of type ``mpg'' (8.5~Mb) and can be found with the \doi{10.1121/1.4939490}.}
\vspace{1cm}
\end{minipage}

\begin{figure}
	\centering
	\includegraphics[width=.65\textwidth]{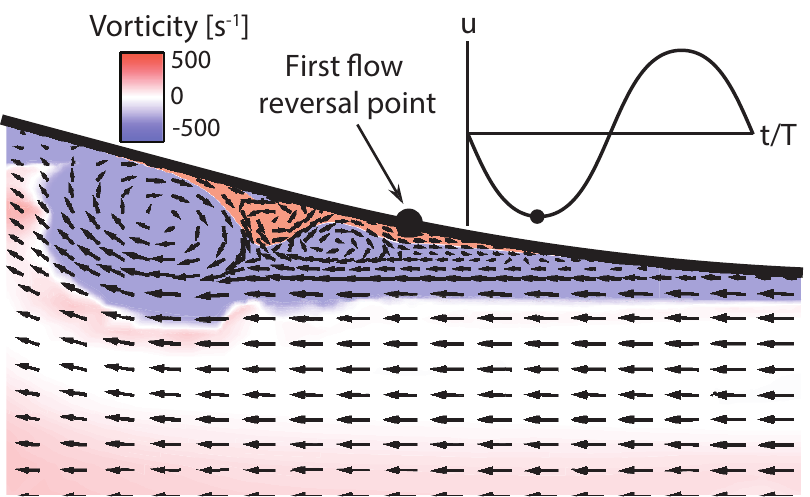}
	\caption{Vector plot of the velocity field in the vicinity of the jet pump waist for the geometry with $\Delta L_{\text{JP}}$ = \SI{10}{\mm}, where $t/T = 0.25$ and $KC_D = 7.4$. The underlaying shades depict the vorticity field.}
	\label{fig:vector_Rc81}
\end{figure}

The spatial point along the wall where flow reversal first occurs is found to shift away from the jet pump waist for an increasing transition length. For all cases, the leftward wake formation caused by the reversed flow was found to begin in the immediate vicinity of this point. An example of this is shown in Fig. \ref{fig:vector_Rc81}, for the case where $\Delta L_{\text{JP}}$ = \SI{10} {\mm} and $t/T = 0.25$. Here a leftward propagating wake is formed from the point of first flow reversal, while the flow to the right of this point remains calm and does not generate any significant minor losses.  

\begin{figure}
	\centering
	\includegraphics[width=.5\textwidth]{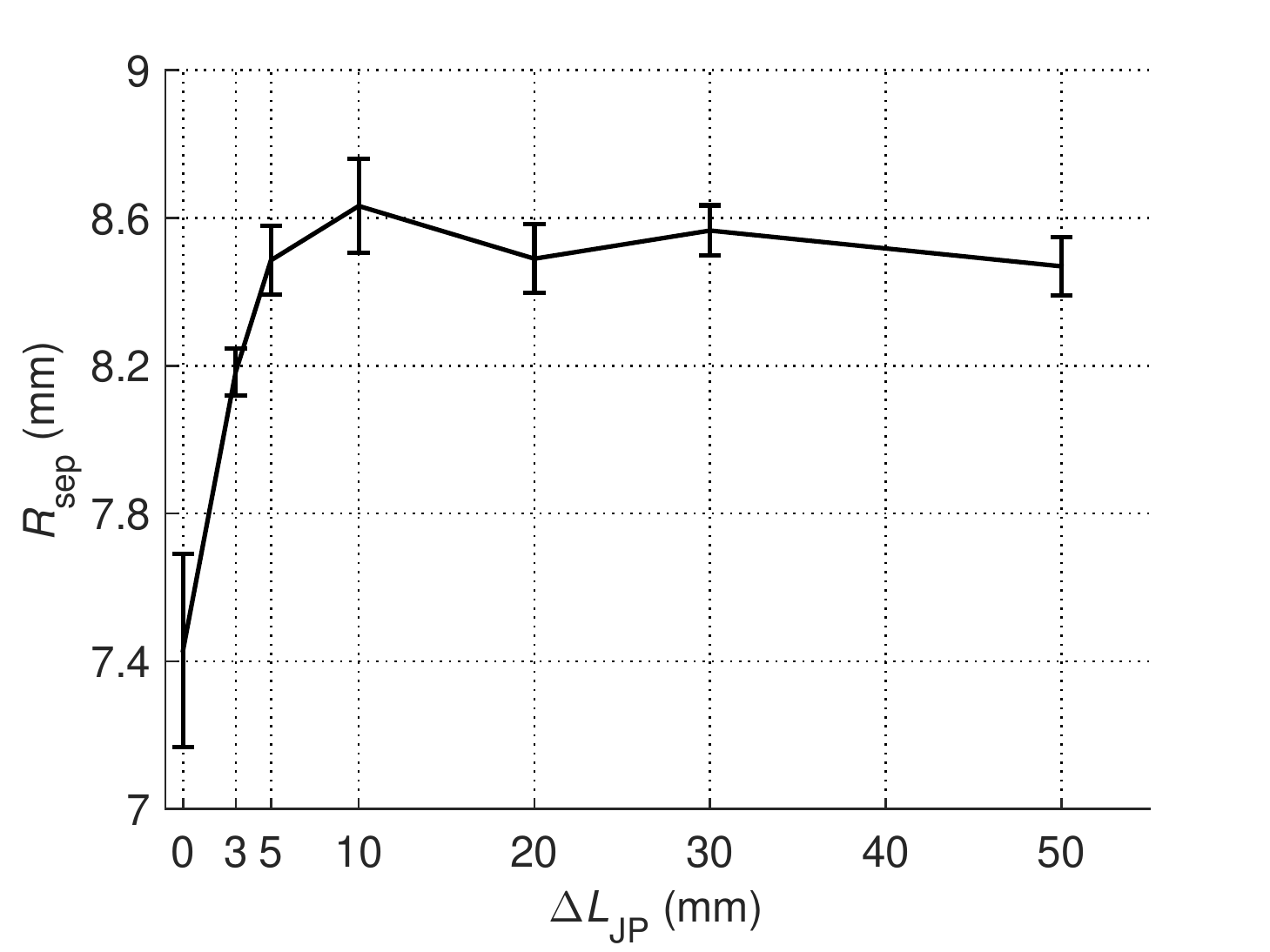}
	\caption{Local radius $R_{\text{sep}}$ at which flow reversal is first found as a function of $\Delta L_{\text{JP}}$. Error bars indicate the standard deviation of $R_{\text{sep}}$ for different displacement amplitudes at a given $\Delta L_{\text{JP}}$.}
	\label{fig:radius_reversal}
\end{figure}

As previously mentioned, the point of first flow reversal is shifted away from the jet pump waist. However, due to the different slopes of the geometries in the vicinity of the jet pump waist, it is more meaningful to look at the local radius $R_{\text{sep}}$ at which flow reversal first occurs. For a specific displacement amplitude and frequency, the local radius will determine the local fluid velocity. This will in turn determine the magnitude of the minor losses of the leftward wake formation. Fig. \ref{fig:radius_reversal} depicts $R_{\text{sep}}$ as a function of $\Delta L_{\text{JP}}$, where the error bars indicate the standard deviation of $R_{\text{sep}}$ for different displacement amplitudes at a given $\Delta L_{\text{JP}}$. For the reference case $R_{\text{sep}}$ = \SI{7.4}{\mm}, which is very close to the radius at the jet pump waist ($R_s$ = \SI{7.0}{\mm}), and hence large minor losses are generated. By elongating the transition length, the local radius of flow reversal can be increased to $R_{\text{sep}} \approx$ \SI{8.5}{\mm} for $\Delta L_{\text{JP}}$ = \SI{5}{\mm}. Further increasing the transition length has no significant effect on $R_{\text{sep}}$ for the geometries that are investigated in this work. Nevertheless, the increase of $R_{\text{sep}}$ will reduce the minor losses generated by the wake formation when compared with the reference geometry, therewith increasing the jet pump performance.     

\subsection{Compact designs}
\label{sec:cut_cases}
In Sec. \ref{sec:increased_trans_length} it is shown that by increasing the transition length between the small jet pump opening and the tapered surface, both the effectiveness and robustness of the jet pump are significantly increased. The downfall of improving the performance in this manner is the increased jet pump length, which prohibits the design of compact thermoacoustic devices. So far we have investigated designs that have a tapered surface up to the outer tube radius (Fig. \ref{fig:geom_rc2}), while jet pumps can be made more compact by cutting the tapered surface at a specific radius $R_b$ (as shown in Fig. \ref{fig:jetpumpgeom}). According to the quasi-steady approximation, the minor losses generated at the big opening will decrease the jet pump performance. However, they are estimated to be a factor $(R_s/R_b)^4$ less than those at the small opening. Therefore, the extent to which the radius of the big opening, and therewith the total jet pump length, can be reduced before a significant performance drop occurs is investigated. This is done for the geometry with $\Delta L_{\text{JP}}$ = \SI{20}{\mm}, as this design is not much longer than the reference geometry but still shows a great performance improvement (see Fig. \ref{fig:transition_length_dp_de}). The total length of the geometry with $\Delta L_{\text{JP}}$ = \SI{20}{\mm} is $L_{\text{JP}}$ = \SI{111.5}{\mm}. The performance of six geometries with a reduced length are compared with this geometry, namely: $L_{\text{JP}}$ = \SI{63.5}{\mm}, \SI{49.5}{\mm}, \SI{45.5}{\mm}, \SI{40.5}{\mm}, \SI{35.5}{\mm}, and \SI{25.5}{\mm}. The corresponding big hole radii are $R_b$ = \SI{17.1}{\mm}, \SI{13.4}{\mm}, \SI{12.3}{\mm}, \SI{11.1}{\mm}, \SI{10.0}{\mm}, and \SI{8.3}{\mm}, respectively. Note that the radius of the small opening, $R_s$, is \SI{7}{\mm} for all geometries. 
     
\begin{figure}
	\centering
	\includegraphics[width=.5\textwidth]{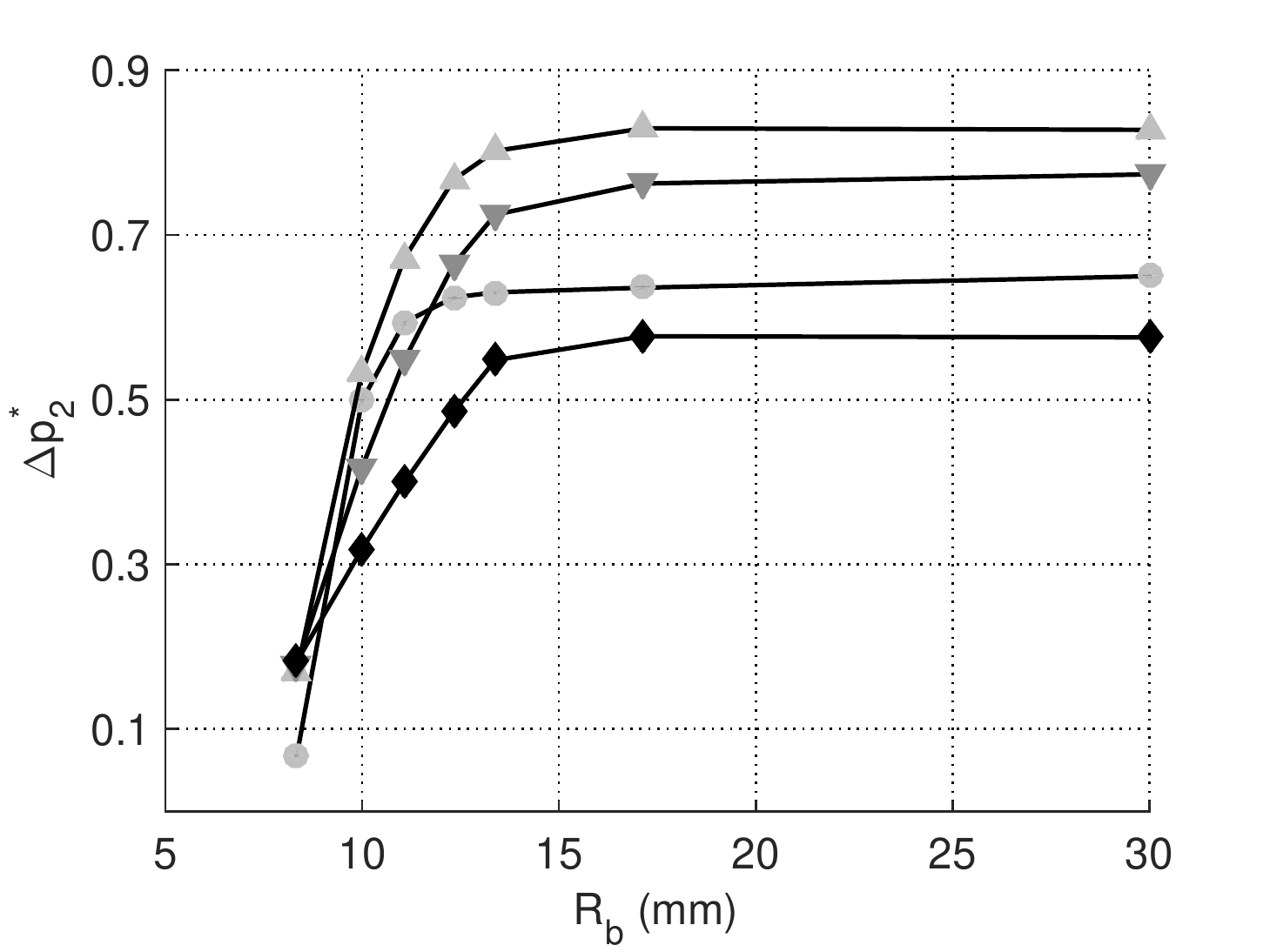}
	\caption{Dimensionless pressure drop as a function of $R_b$. Lines indicate a constant $KC_D$ for the values: $KC_D = 1.3$ ({\Large {\color[rgb]{0.75,0.75,0.75}{$\bullet$}}}),  $KC_D = 2.8$ ({\color[rgb]{0.75,0.75,0.75}{$\blacktriangle$}}), $KC_D = 3.6$ ({\color[rgb]{0.55,0.55,0.55}{$\blacktriangledown$}}), and $KC_D = 5.4$ ({\color[rgb]{0.0,0.0,0.0}{$\blacklozenge$}}).}
	\label{fig:reduced_length_dp_de}
\end{figure}

In Fig. \ref{fig:reduced_length_dp_de} the dimensionless pressure drop is shown as a function of $R_b$ for several Keulegan-Carpenter numbers. The jet pump performance is depicted for values of $KC_D$ in the no flow separation regime ($KC_D = 1.3$ and $KC_D = 2.8$), the partial flow separation regime ($KC_D = 3.6$), and the full flow separation regime ($KC_D = 5.4$). It can be seen that there is no difference in the trend for the various values of $KC_D$. Therefore, reducing the big hole radius has a similar effect in each flow regime. At all Keulegan-Carpenter numbers, no performance drop is present for a small reduction of the big hole radius. For $R_b \lesssim$ \SI{13}{\mm}, the minor losses generated at the big hole become too large, resulting in a significant reduction of $\Delta p^*_2$ (see Fig. \ref{fig:reduced_length_dp_de}) and increase of $\Delta \dot{E}^*_2$. This transition was found at the same big hole radius for the reference geometry ($\Delta L_{\text{JP}}$ = \SI{0}{\mm}), and can therefore be used as a general result for all geometries presented in this work. 

Despite the reduced performance for a geometry with a big hole radius that approaches the small hole radius, the length of jet pumps can still be significantly decreased. As an example, for the cases with $\Delta L_{\text{JP}}$ = \SI{20}{\mm} (Fig. \ref{fig:reduced_length_dp_de}), the geometry with $L_{\text{JP}}$ = \SI{49.5}{\mm} has the same performance as the geometry with $L_{\text{JP}}$ = \SI{111.5}{\mm}. This shows that the jet pumps, as proposed in section \ref{sec:increased_trans_length}, can be made much more compact without reducing their performance.

\subsection{Experimental validation}
\label{sec:exp_validation}
To validate the numerical results presented in this work, the jet pump design with $\Delta L_{\text{JP}}$ = \SI{20}{\mm}, $R_b$ = \SI{13.4}{\mm}, and $L_{\text{JP}}$ = \SI{49.5}{\mm} is constructed using a 3D printing method. The performance of the jet pump is investigated in an experimental set-up that is similar to the one previously used by Aben. \cite{Aben2010} The jet pump is situated in a tube with two pressure sensors on either side of the jet pump. An acoustic wave with a frequency of \SI{80}{\Hz} is generated by a loudspeaker. At each displacement amplitude, the mean pressure and the pressure amplitude are determined by averaging over \SI{60}{\s}. For accuracy, the data where the standard deviation of the pressure is larger than 10\% of the mean pressure has been left out. From the remaining pressure data in the laminar regime, the values for $KC_D$, $\Delta p^*_2$, and $\Delta \dot{E}^*_2$ are calculated. 

\begin{figure}
	\centering
	\includegraphics[width=.5\textwidth]{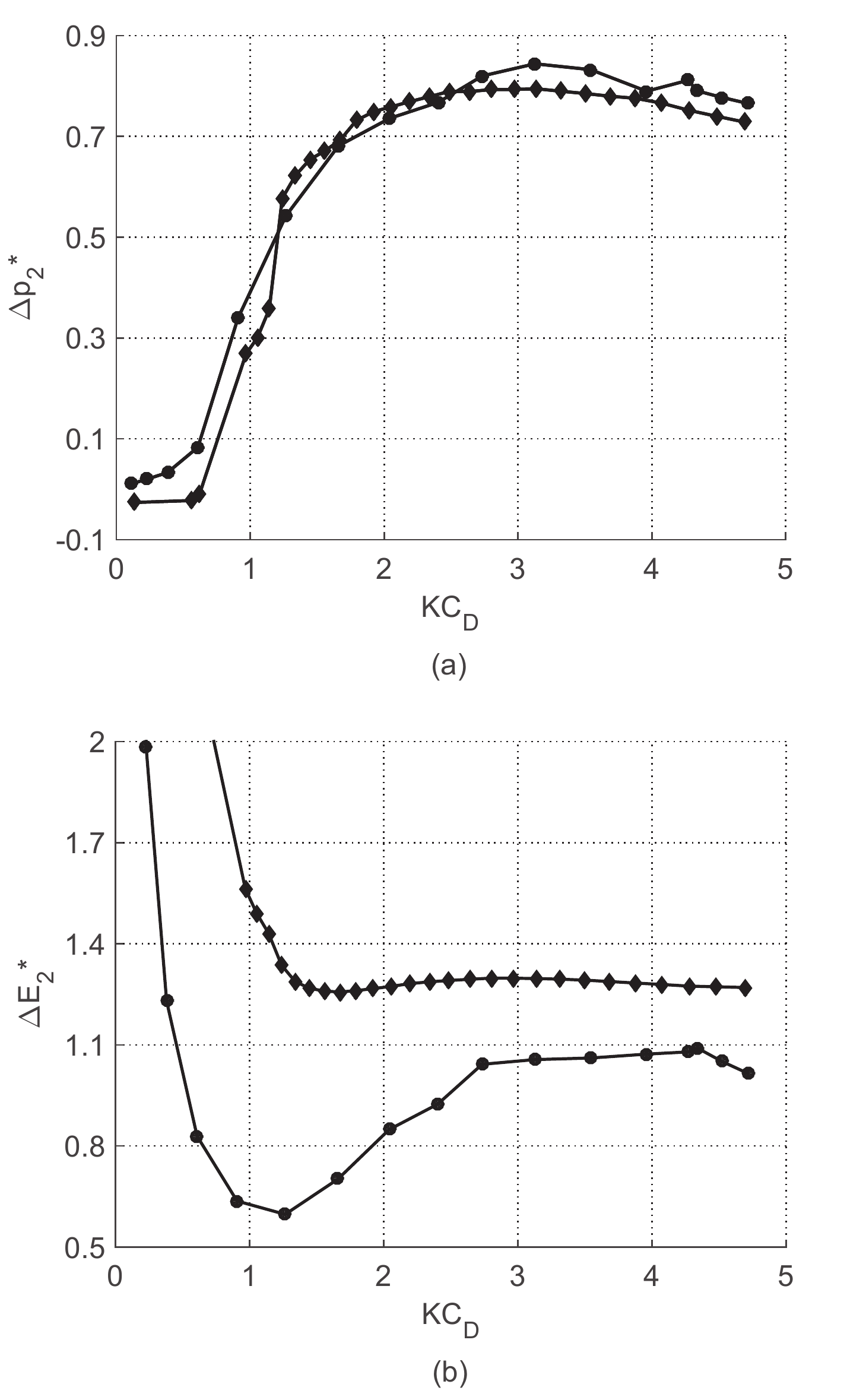}
	\caption{Dimensionless pressure drop (a) and acoustic power dissipation (b) as a function of $KC_D$ for the numerical simulations ({\Large $ \bullet $}) and experimental investigation ($\blacklozenge$) at 80 Hz. Experimental data where the standard deviation of the pressure is larger than 10\% of the mean pressure has been omitted.}
	\label{fig:experimental_dp}
\end{figure}

Fig. \ref{fig:experimental_dp} shows the dimensionless pressure drop (a) and acoustic power dissipation (b) for the experimental and numerical results as a function of $KC_D$. For all values of $KC_D$, there is a good correspondence between the experimentally and numerically determined dimensionless pressure drop. The initial increase, the maximum, and the following decrease of $\Delta p^*_2$ are all very similar. This shows that the numerical model is able to accurately predict the dimensionless pressure drop. 

The dimensionless acoustic power dissipation (Fig. \ref{fig:experimental_dp}(b)) for the experiments is significantly higher compared with the numerical results for the entire range of $KC_D$. However, the trends in $\Delta \dot{E}^*_2$ are found to be the same. In recent experiments the effect of the surface roughness on the jet pump performance has been investigated, and it is found that it has a significant effect on $\Delta \dot{E}^*_2$. The difference is large enough to explain the discrepancy between the experiments with the 3D printed jet pump and the numerical results, where no surface roughness is modeled. This, combined with the good prediction of the dimensionless pressure drop (Fig. \ref{fig:experimental_dp}(a)), establishes enough confidence in the validity of the numerical results presented in this work. Furthermore, the good correspondence shows that the proposed jet pump designs with an increased transition length not only have an improved performance numerically, but also experimentally.

\section{Conclusions}
A computational fluid dynamics model is used to investigate the separation of oscillatory flows in jet pumps. A decrease in jet pump performance is shown to be the direct result of separation in the diverging flow direction. The time in the period at which flow reversal first occurs reduces for an increasing displacement amplitude. This further degrades the jet pump performance by both a decreasing $\Delta p^*_2$ and increasing $\Delta \dot{E}^*_2$.

By increasing the transition length between the small jet pump opening and the tapered surface, the onset of flow separation is shifted to higher displacement amplitudes compared with the reference geometry. This results in an increase of the maximum dimensionless pressure drop from $\Delta p^*_2 = 0.66$ for the reference geometry up to $\Delta p^*_2 = 0.91$ for an increased transition length of $\Delta L_{\text{JP}}$ = \SI{50}{\mm}. Furthermore, the time at which flow reversal first occurs is delayed and the first flow reversal point is shifted to a location with a larger local radius. This reduces the minor losses of the leftward wake formation in the flow separation regime, therewith increasing $\Delta p^*_2$ and reducing $\Delta \dot{E}^*_2$. Increasing the transition length of the reference geometry therefore ensures a more robust and effective jet pump design. 

The designs with an increased transition length have more acoustic power dissipation in the regime without flow separation due to their elongated geometry. Next to that, the increased length prohibits the design of compact thermoacoustic devices. Therefore, it is shown that these jet pump designs can be made more compact by reducing the radius of the big opening without degrading the jet pump performance. To further increase the compactness, jet pumps with multiple holes or a larger taper angle should be investigated. 

The results of the numerical model correspond well with the experimental investigation. Nevertheless, all simulations are performed in the laminar regime, since no suitable turbulence model for oscillatory flows exists. To include the effect that turbulence will have on flow separation and the jet pump performance, future work should focus on developing and validating a relatively low-cost turbulence model that can be used in turbulent oscillatory flow. Alternatively, the reduction of flow separation and its effect on the jet pump performance can be investigated experimentally in the turbulent regime.

\medskip

\noindent \textbf{Acknowledgements}

\setlength{\parindent}{0.7cm} 
Bosch Thermotechnology and Agentschap~NL are thankfully acknowledged for the financial support as part of the EOS-KTO research program under project number KTOT03009.

\clearpage

\end{document}